\RequirePackage{fix-cm}
\documentclass{article}
\usepackage{hyperref}
\usepackage[left=2.0cm,right=2.0cm,top=2.5cm,bottom=2.5cm]{geometry}
\usepackage{color}
\usepackage{graphicx}
\usepackage{amsmath,amssymb}
\usepackage{caption}
\usepackage{subcaption}
\title{Nonreciprocal phase shifts in a nonlinear periodic waveguide  
}

\author{Ali Kogani\footnotemark[1] 
\and Behrooz Yousefzadeh\footnotemark[2]}

\begin{document}
\renewcommand{\thefootnote}{\fnsymbol{footnote}}
\footnotetext[1]{Department of Mechanical, Industrial \& Aerospace Engineering, Concordia University, Montreal, QC, H3G1M8, Canada

  (\href{mailto:ali.kogani@mail.concordia.ca}{ali.kogani@mail.concordia.ca})
}
\footnotetext[2]{Department of Mechanical, Industrial \& Aerospace Engineering, Concordia University, Montreal, QC, H3G1M8, Canada

  (\href{mailto:behrooz.yousefzadeh@concordia.ca}{behrooz.yousefzadeh@concordia.ca})}

\maketitle

\begin{abstract}
We explore nonreciprocal vibration transmission in a nonlinear periodic waveguide. Nonlinearity and asymmetry, the two necessary requirements for nonreciprocity, are both introduced within the unit cell of the periodic waveguide. We focus primarily on the contribution of phase to the nonreciprocal steady-state response of the system. To highlight the phase effects, which are rarely discussed in the literature, we investigate response regimes in which nonreciprocity is solely due to \emph{nonreciprocal phase shifts}: when the locations of the source and receiver are interchanged, the amplitude of transmitted vibrations remains unchanged but the transmitted phases are not equal. We present a computational analysis of this state of \emph{phase nonreciprocity} in the weakly nonlinear frequency-preserving response regime, where we characterize the response using its nonreciprocal phase shift. This allows us to systematically find a set of system parameters (including two symmetry-breaking parameters) that lead to reciprocal nonlinear response in a system with broken mirror symmetry. In other words, we show that breaking the mirror symmetry of a passive nonlinear waveguide is a necessary but insufficient condition for nonreciprocal dynamics to exist. Our findings highlight the important role of phase in nonlinear nonreciprocity and showcase the potential of asymmetry to serve as an additional design parameter.

\end{abstract}

\section{Introduction}
\label{intro}

Reciprocity refers to a symmetry property of a vibrating system that dictates that transmitted vibrations from point $A$ to point $B$ are identical in amplitude, frequency and phase to those from point $B$ to point $A$. In other words, the transmission characteristics between two points do not depend on the direction. This property has been studied and utilized in applications since the nineteenth century, with contributions from Helmholtz~\cite{von1896theorie} and Rayleigh~\cite{strutt_general_1871} among others. In addition to its numerous theoretical implications~\cite{achenbach_reciprocity_2003}, reciprocity has enabled various experimental approaches, for example in vibroacoustics~\cite{fahy_applications_2003,ten_wolde_reciprocity_2010}, structural dynamics~\cite{ewins_modal_2009,TPA}, defect detection and determination of elastic constants~\cite{auld_general_1979}, ultrasonics~\cite{anderson_use_1998} and seismology and geophysics~\cite{knopoff_seismic_1959}. 

There exist several ways for the response of a mechanical system to exhibit nonreciprocal characteristics~\cite{nassar_nonreciprocity_2020}. The presence of nonlinear forces within the structure is a common way to realize nonreciprocity in a passive way, {\it i.e.}, with no need for external bias or activation. In this work, we focus exclusively on this approach. Comprehensive discussion of other approaches to realize nonreciprocity are found elsewhere~\cite{nassar_nonreciprocity_2020,caloz_electromagnetic_2018,reiskarimian_nonreciprocal_2019}. 

Among the appealing features of nonlinear nonreciprocity is that it makes the transmission characteristics between two points dependent on the direction of wave propagation. Thus, nonreciprocity could provide control over the direction of wave propagation, enabling or inhibiting wave transmission in specific directions. For instance, Liang et al. proposed an acoustic diode by combining a superlattice with a highly nonlinear medium, demonstrating a pronounced rectifying effect in the energy flux within a specific frequency range~\cite{liang_acoustic_2009}. Boechler et al. employed a defective granular chain composed of spherical beads to achieve nonreciprocal transmission thresholds for harmonic excitation within a bandgap~\cite{boechler_bifurcation-based_2011}. The response of a nonlinear system is generally dependent on the energy of the system (amplitude of motion), but there do exist response regimes that are independent of the energy level as well. For example, Lu and Norris used bilinear stiffness to realize a diode-like waveguide, in which waves with distinct patterns are transmitted in different directions regardless of the wave amplitude~\cite{lu_non-reciprocal_2020}.

The existence of nonlinear forces is not a sufficient condition for nonreciprocal response to exist. For example, the response of a system or transmission channel that possesses mirror symmetry remains reciprocal even in the presence of strong nonlinearity: the right-to-left transmission path is identical to the left-to-right path by virtue of symmetry. Thus, it is necessary to break the mirror symmetry of the system to enable a nonreciprocal response~\cite{lepri_asymmetric_2011}. In periodic systems, which are our focus, nonlinearity and asymmetry can be introduced at the level of the unit cell, thereby appearing periodically throughout the structure; see~\cite{luo_non-reciprocal_2018, fang_broadband_2020, moore_nonreciprocity_2018, brandenbourger_non-reciprocal_2019} for examples implementing this approach. Alternatively, nonlinear nonreciprocity is achieved by incorporating asymmetry either locally (such as a gate or defect)~\cite{liang_acoustic_2009, boechler_bifurcation-based_2011,darabi_broadband_2019,wang_non-reciprocal_2023,gatePai} or gradually throughout the system~\cite{lu_non-reciprocal_2020,tapered}.

The focus of studies on nonlinear nonreciprocity is on maximizing the nonreciprocal energy transfer: the difference in the transmitted energy or amplitude when the locations of the source and receiver are exchanged. Within this context, a highly nonreciprocal response corresponds to unidirectional transmission of energy within a system, which enables targeted energy transfer~\cite{TET}. In this work, we take on a different perspective to address a question that is primarily overlooked: how does the difference in the transmitted phases (not energies) contribute to nonreciprocity? 

To understand the influence of phase, we investigate response regimes that are characterized by nonreciprocal phase shifts but equal transmitted energies: when the locations of the source and receiver are interchanged, the amplitude of transmitted vibrations remains unchanged but the transmitted phases are not equal. We refer to this response regime as {\it phase nonreciprocity} because a phase difference between the forward (left-to-right) and backward (left-to-right) configurations is the only contributor to nonreciprocity. 

Phase nonreciprocity is already reported in systems with two degrees of freedom that have well-separated modes~\cite{yousefzadeh_computation_2022}. We extend and expand this work to periodic waveguides, where there exists modal overlap within a pass band. We perform a parametric study to investigate the influence of forcing amplitude, damping and asymmetry on phase nonreciprocity. It is already known that asymmetry is a necessary but insufficient condition for realizing nonreciprocity in nonlinear systems~\cite{giraldo_restoring_2023,recip01,blanchard_non-reciprocity_2018}. In the context of phase nonreciprocity, a zero nonreciprocal phase shift corresponds to a reciprocal response. We thereby utilize this approach to systematically find parameters that enable a non-trivial reciprocal response in a system with broken mirror symmetry. 

We focus exclusively on the steady-state response of a nonlinear periodic system to external harmonic excitation. Because superposition does not hold for nonlinear systems, the findings will not identically extend to the response of the system to other types of excitation. See~\cite{PhysRevApplied.15.034005, PhysRevApplied.12.034033, fronk_acoustic_2019} for examples of recent studies of nonlinear nonreciprocity in systems subject to impulse excitation. 

We perform our study for a periodic system that comprises eight unit cells, each with two degrees of freedom. Nonlinearity and asymmetry are both introduced within the unit cell. The source of nonlinearity is the cubic restoring force that grounds every mass in the system. The ratio of the two masses and the two springs in the unit cell act as two symmetry-breaking parameters in this study. We compute the locus of phase nonreciprocity and characterize the response using the nonreciprocal phase shift. 

Section~\ref{sec:1} introduces the details of the setup and solution methodology. We discuss the frequency response curves of the system in Section~\ref{frf} and introduce phase nonreciprocity. We present the results of our parametric study of phase nonreciprocity in Section~\ref{DeltaPhi}, where we use the mass ratio as the symmetry-breaking parameter. In Section~\ref{restoring}, we discuss how to use the stiffness ratio to restore reciprocity even in the absence of mirror symmetry. We conclude in Section~\ref{conclusion} by an overview of our findings. 

\section{Problem Setup}
\label{sec:1}

\begin{figure}

  \includegraphics[width=1\textwidth]{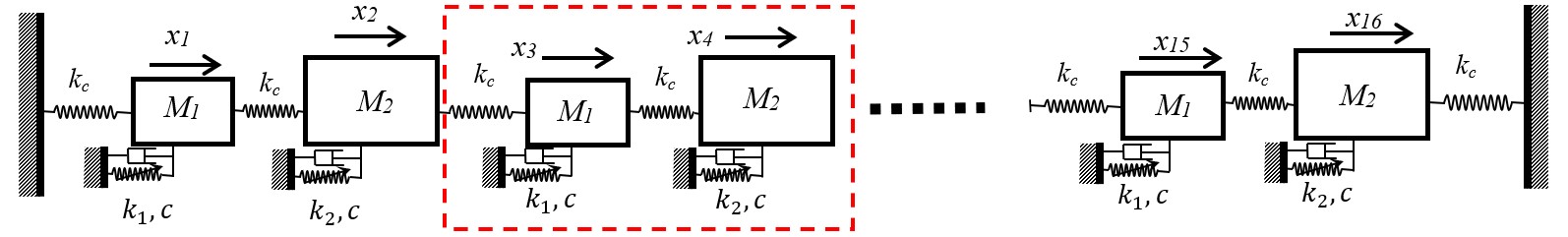}
\caption{Schematic of the finite periodic structure. The red dashed box shows the unit cell, which consists of two linear oscillators coupled with a linear stiffness and grounded by a nonlinear spring and a linear damping. }
\label{fig:1}      
\end{figure}

Fig.~\ref{fig:1} shows a schematic representation of the periodic system we study in this work. The structure comprises eight identical unit cells, with fixed boundaries at the two ends. Each cell consists of two masses, $M_1$ and $M_2=\mu M_1$, which are coupled by a linear spring $k_c$. The mass $M_1$ is anchored to the ground by a spring with cubic nonlinearity, $k_1=k_g+k_n\delta^2$, where $\delta$ represents the spring deformation from the static equilibrium position. The mass $M_2$ is anchored to the ground with a similar nonlinear spring of constant $k_2=k^\prime_g+k_n\delta^2$. The linear components of the two grounding springs are different, but the coefficient of the cubic term is the same for the two springs. Adjacent unit cells are coupled by the same linear spring that couples $M_1$ and $M_2$ within each cell. Energy loss is implemented in the model by identical linear viscous dampers of constant $c$ that connect each mass to the ground. The periodic system is subject to harmonic external force of amplitude $F$ and frequency $\omega_f$ (not shown). 

The mirror symmetry of the system can be broken within a unit cell by changing the ratio of the two masses, $\mu=M_2/M_1$, or by changing the ratio of the two grounding springs, $r=k^\prime_g/k_g$.

\subsection{Governing equations}
\label{EOM}

As outlined in Appendix A, the equations of motion for the system shown in Fig.~\ref{fig:1} can be written using non-dimensional parameters as 
\begin{equation}
\label{govern}
\begin{aligned}
\bar{x}''_{2i-1}+2K_c\bar{x}_{2i-1}-K_c(\bar{x}_{2i-2}+\bar{x}_{2i})+\bar{x}_{2i-1}+K_n\bar{x}_{2i-1}^3+2\zeta_g\bar{x}'_{2i-1}=F_{2i-1}\cos{\omega_f\tau} \\
\mu\bar{x}''_{2i}+2K_c\bar{x}_{2i}-K_c(\bar{x}_{2i+1}+\bar{x}_{2i-1})+r_g\bar{x}_{2i}+K_n\bar{x}_{2i}^3+2\zeta_g\bar{x}'_{2i}=F_{2i}\cos{\omega_f\tau}
\end{aligned}
\end{equation}
The subscripts $i=1,...,8$ denote the counter of unit cells. We use $\bar{x}_0(t)=0$ and $\bar{x}_{17}(t)=0$ to represent the fixed boundaries at the two ends of the structure. With the exception of $F_1$ and $F_{16}$, all the forcing amplitudes are zero. 

Testing for reciprocity requires comparing the response of the system for two different configurations of the input-output locations. We study the end-to-end transmission characteristics of the finite periodic system in this work. Accordingly, we define (i) the {\it forward} configuration with $F_1=P$ and $F_{16}=0$, in which the output is the displacement of the last (right-most) mass, $\bar{x}_{16}^F(t)$; (ii) the {\it backward} configuration with $F_1=0$ and $F_{16}=P$, in which the output is the displacement of the first (left-most) mass, $\bar{x}_1^B(t)$. The response of the system is reciprocal if and only if $\bar{x}_{16}^F(t)=\bar{x}_1^B(t)$.

\subsection{Methodology}

Systems with cubic nonlinear elasticity, also known as Duffing oscillators, have long been known to exhibit a rich nonlinear behavior and bifurcation structure~\cite{guckenheimer_nonlinear_2013,duffing}. In this work, we exclusively focus on the weakly nonlinear response regime of the system in Fig.~\ref{fig:1}, described by Eq.~(\ref{govern}). In particular, the steady-state response of the system remains harmonic in this regime and the only bifurcations encountered are the saddle-node bifurcation of periodic orbits at the turning points of the frequency response curves. These points are characterized visually by the vertical tangencies of the frequency response curves or, more precisely, to points at which one of the Floquet multipliers of the system exits the unit circle on the positive real axis~\cite{guckenheimer_nonlinear_2013}. 

In this light, we adopt a harmonic representation for the steady-state response of Eq.~(\ref{EOM}) as 
\begin{equation}
    \label{eqH}
    \bar{x}_j(t)={X}_j\cos{(\omega_ft+\phi_j)}
\end{equation}
for $j=1, ... ,16$. The response is harmonic with its frequency imposed by the external force but different amplitude and phase from those of the external excitation. 
Eq.~(\ref{eqH}) is then substituted in Eq.~(\ref{govern}) to convert the governing equations from a set of nonlinear ordinary differential equations to a set of nonlinear algebraic equations for the unknown amplitudes, $X_j$, and phases, $\phi_j$. The steady-state response of the system is obtained by solving the ensuing set of algebraic equations; we use the Matlab package {\sc coco} for this purpose~\cite{coco}. The linear stability analysis of the response is determined by computing the Floquet multipliers of the system~\cite{vonGroll}. 
Detailed formulation of the numerical technique, the harmonic balance method, along with descriptions of its more sophisticated implementations can be found elsewhere~\cite{detroux_harmonic_2015,krack_harmonic_2019}. 

For ease of future reference, we define the output displacements for the forward and backward configurations, $x_{16}^F(t)$ and $x_1^B(t)$, as follows
\begin{subequations}
    \begin{align}
        \bar{x}_{16}^F(t) &= A^F\cos(\omega_f t + \phi^F)\\ 
        \bar{x}_1^B(t)    &= A^B\cos(\omega_f t + \phi^B)
    \end{align}    
\end{subequations}
We refer to $A^{F,B}$ and $\phi^{F,B}$ as the output or transmitted amplitude and phase of the forward or backward configurations.

\subsection{Response norms}
\label{norms}

To quantify the response of the system, we define the output norms for the forward ($N^F$) and backward ($N^B$) configurations as 
\begin{subequations}
\label{Norms}
    \begin{align}
    N^F &= \frac{1}{T}\int_{0}^{T}{\left(\bar{x}_{16}^F\left(t\right)\right)^2dt}\\
    N^B &= \frac{1}{T}\int_{0}^{T}{\left(\bar{x}_1^B\left(t\right)\right)^2dt}
    \end{align}
\end{subequations}
where $T=2\pi/\omega_f$ is the period of excitation. For a harmonic response, we have $N^F=|X_{16}^F|^2/2$ and $N^B=|X_1^B|^2/2$. Even though the response of the system remains harmonic in the parameter ranges discussed in this work, we continue using the output norms in Eq.~(\ref{Norms}) instead of response amplitudes. This keeps our approach consistent with our previous work~\cite{kogani_nonreciprocal_2022,yousefzadeh_computation_2022,giraldo_restoring_2023}.

To quantify the degree of nonreciprocity in the response, we define the reciprocity bias as
\begin{equation} 
    \label{Eq:Reciprocity}
    R=\frac{1}{T}\int_{0}^{T}({\bar{x}_{16}^F\left(t\right)-\bar{x}_1^B\left(t\right))}^2dt
\end{equation}
The response of the system is reciprocal if and only if $R=0$.
When needed, we normalize the reciprocity bias with the output norms, 
\begin{equation}
    \label{RN}
    R_N=\frac{R}{N^B+N^F}
\end{equation}
to eliminate the apparent increase in reciprocity that is caused merely by higher response amplitudes.

\section{Frequency Response Curves}
\label{frf}

We start by studying the frequency response curves of the system: the output norms, $N^F$ and $N^B$, as a function of the forcing frequency, $\omega_f$. We keep moderate coupling, $K_c=1$, throughout the work to avoid the overlapping of the two passbands of the system. The nonlinear stiffness is of the hardening type, $K_n=1$. We consider moderate damping, $\zeta_g=0.03$, except in section~\ref{damping} where we explore the role of damping on phase nonreciprocity. We choose $P=0.15$ for the forcing amplitude throughout this section; we study the role of $P$ in section~\ref{Fampl}. 

\begin{figure}
  \includegraphics[width=1\textwidth]{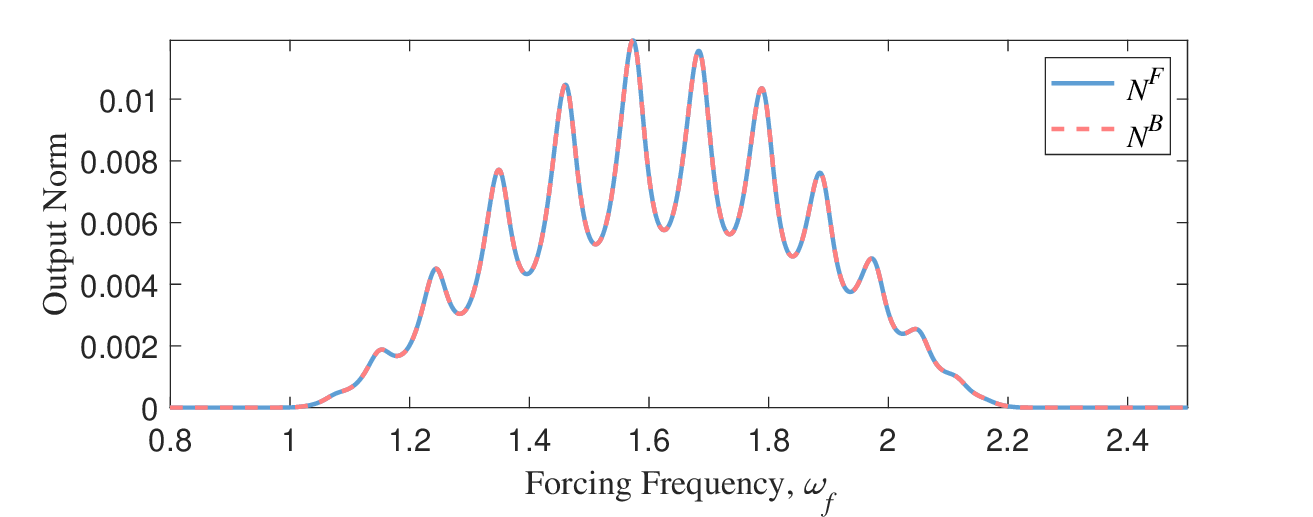}
\caption{Output norms (Eq. \ref{Norms}) at $P=0.15$ for the system with mirror symmetry ($\mu=1$)}
\label{fig:2}      
\end{figure}

To better appreciate the role of symmetry in enabling a nonreciprocal response, we present the response of the mirror-symmetric system in Fig.~\ref{fig:2}. Owing to the mirror symmetry of the system ($\mu=1, ~r=1$), the forward and backward configurations are identical. This corresponds to the identical frequency response curves in Fig.~\ref{fig:2} despite the presence of nonlinearity. Note that the response is nonlinear even though the typical `bends' in the frequency response curves are not present for this particular set of system parameters.

\begin{figure}
  \includegraphics[width=1\textwidth]{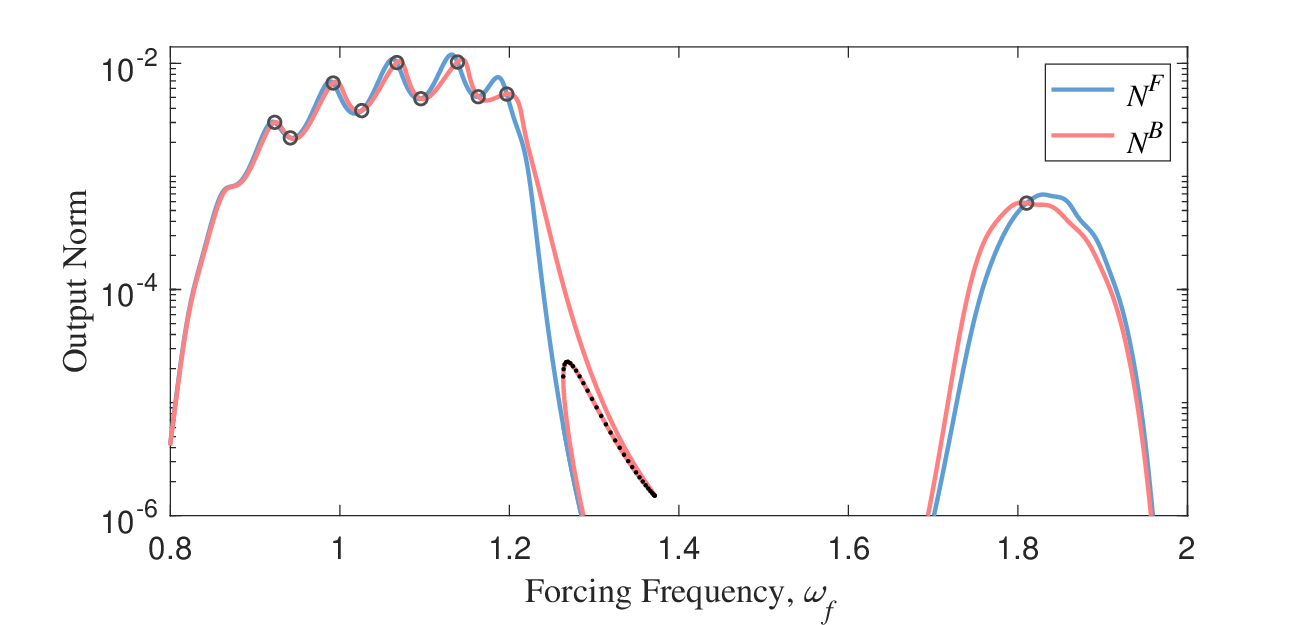}
\caption{Frequency response curves at $P=0.15$ for the system with broken mirror symmetry, $\mu=2$. The grey circles show points at which $N^F=N^B$. The black dots indicate unstable regions in the response.}
\label{fig:3}  
\end{figure}

Fig.~\ref{fig:3} shows the frequency response curves of the system after we break the mirror symmetry of the unit cell by changing the mass ratio, $\mu=2$. Notice that the periodic system with different masses exhibits a band gap where the output norm is very small; roughly for $1.3<\omega_f<1.7$. This is because the unit cell of the lattice has two degrees of freedom when $\mu\ne1$. The lower branch (first pass band) is known as the acoustic (in-phase) branch, which corresponds to the collection of natural frequencies at which the adjacent masses move in phase with each other. The upper branch (second pass band) is known as the optical (out-of-phase) branch, corresponding to mode shapes at which adjacent masses move out of phase with each other. As a result, damping has a larger influence on the optical branch and the amplitudes of the acoustic branch are relatively higher. The resonances of the optical branch are corresponding more highly damped. 

As expected, breaking the mirror symmetry of the system enables the response of the system to become nonreciprocal. This is, of course, most conspicuous near the resonance frequencies because the amplitude of motion is relatively higher. The response away from resonances is linear owing to its small amplitude and necessarily reciprocal. Even near the resonances, however, we notice there are frequencies at which the two frequency response curves intersect ($N^F=N^B$), indicating equal amplitudes in the forward and backward configurations. 

\begin{figure}
\begin{subfigure}{1\textwidth}
  \includegraphics[width=\textwidth]{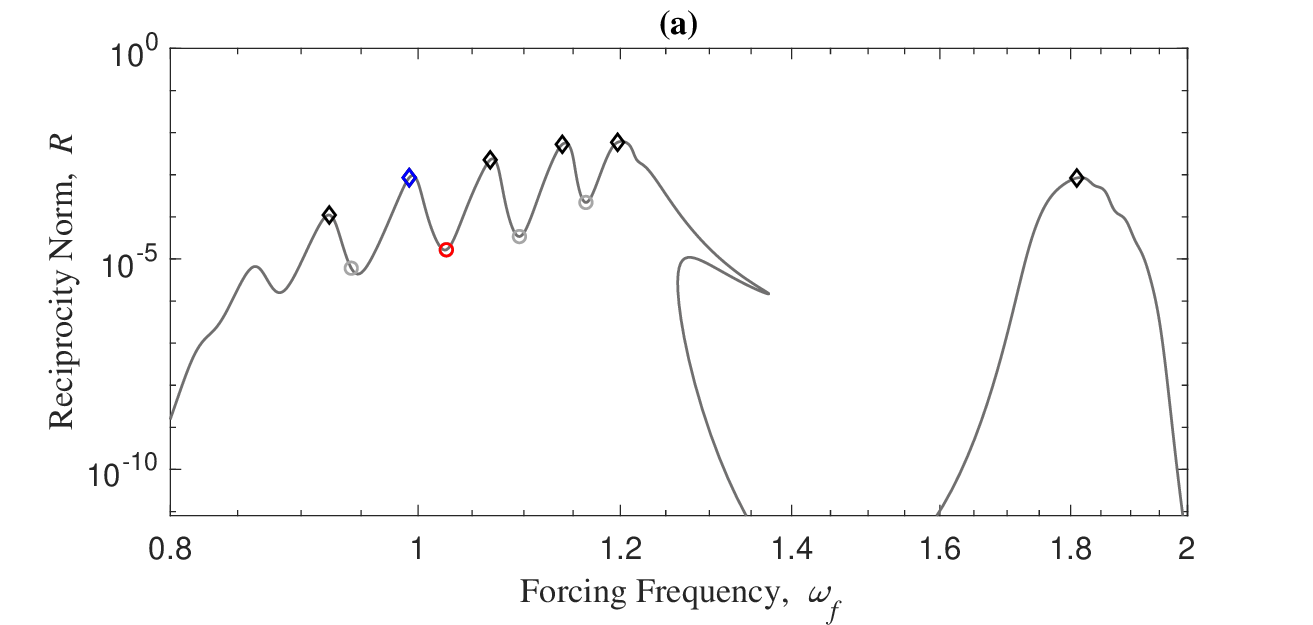}
\end{subfigure}

\begin{subfigure}[b]{.49\textwidth}
\includegraphics[width=\textwidth]{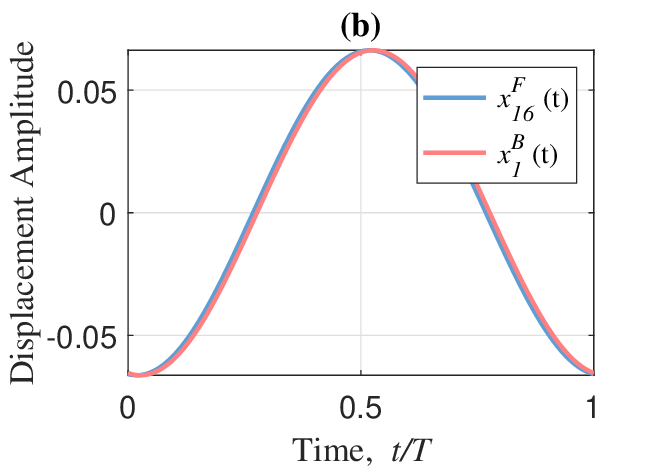}
\end{subfigure}
\hfill
\begin{subfigure}[b]{.49\textwidth}
\includegraphics[width=\textwidth]{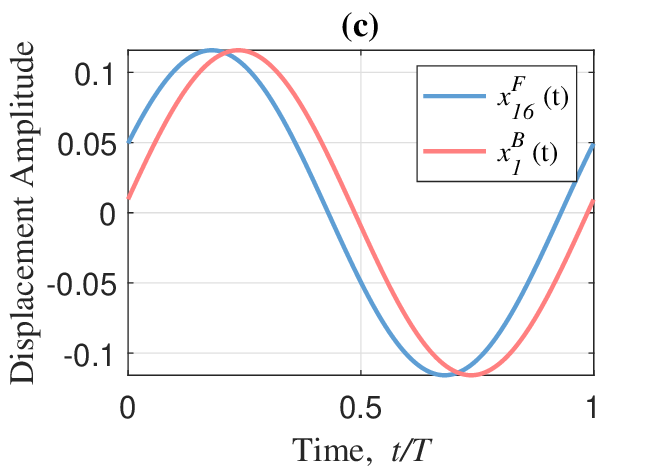}
\end{subfigure}
\caption{Nonreciprocal dynamics for $P=0.15$ and $\mu=2$. (a)~The reciprocity norm, $R$, with the markers indicating $N^F=N^B$. Diamonds denote phase nonreciprocity $(N^F=N^B,~\phi^F\ne\phi^B)$ and circles denote frequencies at which the response of the system is almost reciprocal $(N^F=N^B,~\phi^F\approx\phi^B)$. (b)~The time-domain response at the red circle, $\omega_f=1.025$. (c)~The time-domain response at the blue diamond, $\omega_f=0.99$.
}
\label{fig:4}
\end{figure}

The grey circles in Fig.~\ref{fig:3} indicate the intersection points of the two frequency response curves. Some of these intersections occur near resonances and some near anti-resonances of the system. To understand the difference between these points, we refer to the reciprocity bias, $R$, shown in Fig.~\ref{fig:4}(a). 

Fig.~\ref{fig:4}(a) indicates the points at which the forward and backward transmissions have the same amplitude, $N^F=N^B$, with two different markers. The circle markers in the valleys correspond to frequencies at which the response of the system is almost reciprocal. Fig.~\ref{fig:4}(b) shows the output displacement at the red circle ($\omega_f=1.025$) in the time domain. These points occur near the local minima of the frequency response curves, which correspond to the anti-resonances of the system. Because of the relatively low amplitude of the response, the system behaves almost linearly near these points and almost reciprocally as a result. In practice, these points are not of much importance due to their low response amplitudes. 

The diamond markers in Fig.~\ref{fig:4}(a) correspond to frequencies at which $N^F=N^B$ in the vicinity of the resonances of the nonlinear system. Even though the amplitude of the transmitted vibrations is equal for the forward and backward configurations, these points occur near the local maxima of the reciprocity bias. Fig.~\ref{fig:4}(c) shows the time-domain response at the blue diamond ($\omega_f=0.99$), revealing that the only difference between the forward and backward configurations is in the transmitted phases: $\phi^F\ne\phi^B$. We refer to this state as the state of {\it phase nonreciprocity} and to $\Delta\phi=\phi^F-\phi^B$ as the {\it nonreciprocal phase shift} of the response~\cite{yousefzadeh_computation_2022}. 

Before moving on to a detailed analysis of phase nonreciprocity in section~\ref{DeltaPhi}, we note in Fig.~\ref{fig:4}(a) that the value of $R$ corresponding to the bent portion of the curve ($1.3<\omega_f<1.4$) is markedly smaller than those at the diamond markers. This indicates the significant contribution of nonreciprocal phase shifts to the degree of nonreciprocity. The influence of phase, a key focus of the present work, is typically overlooked in the analysis of nonreciprocity in nonlinear systems. 

\section{Phase Nonreciprocity}
\label{DeltaPhi}

To better understand phase nonreciprocity, we perform a parametric study of the system while imposing the constraint $N^F=N^B$. Specifically, we investigate the effect of the asymmetry of the system as controlled by the mass ratio, $\mu$, the influence of the forcing amplitude, $P$, and damping effects, $\zeta_g$. The mass ratio, $\mu$, remains the only source of asymmetry in this section ($r_g=1$). We postpone discussing the effect of two competing symmetry-breaking parameters to section~\ref{restoring}.

\subsection{Influence of asymmetry (mass ratio, $\mu$)}
\label{mu}

\begin{figure}
\begin{subfigure}[b]{.49\textwidth}
  \includegraphics[width=\textwidth]{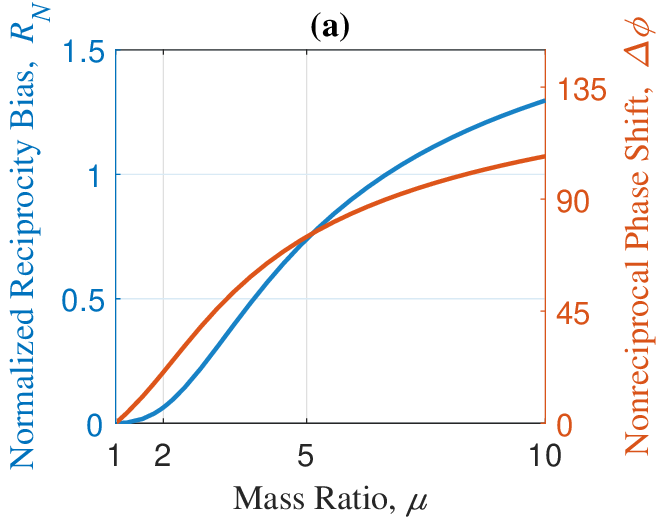}
\end{subfigure}
\hfill
\begin{subfigure}[b]{.49\textwidth}
      \includegraphics[width=\textwidth]{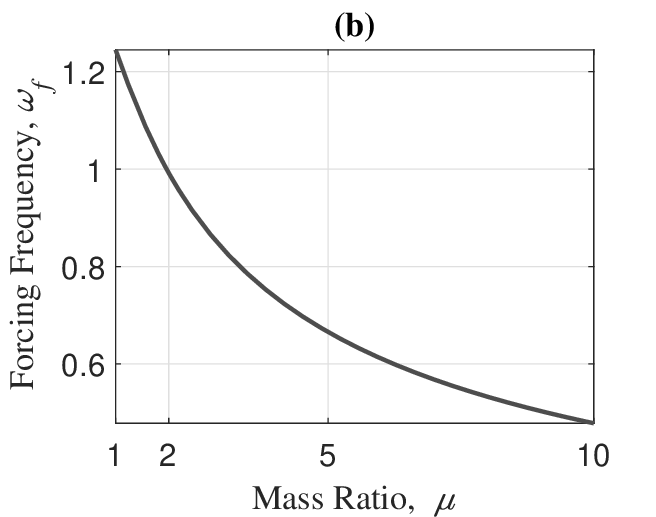}
\end{subfigure}

    \caption{Locus of phase nonreciprocity ($N^F=N^B$) as a function of the asymmetry parameter, $\mu$, at $P=0.15$. (a)~Normalized reciprocity bias, $R_N$, and the phase shift between the forward and backward configurations, $\Delta\phi=\phi^F-\phi^B$. (b)~The forcing frequency, $\omega_f$, at which phase nonreciprocity occurs.}
    \label{fig:5}
\end{figure}

We study the influence of asymmetry on phase nonreciprocity by computing the locus of $N^F=N^B$ as a function of the mass ratio for $1\le\mu\le10$. Fig.~\ref{fig:5} shows this locus for different parameters. Panel (a) shows the variation of the normalized reciprocity bias, $R_N$, and the nonreciprocal phase shift, $\Delta\phi$. As expected, the response is reciprocal at $\mu=1$ because the system is mirror-symmetric at that point. Increasing the asymmetry of the system results in an increase in the degree of nonreciprocity, as indicated by the increasing value of $R_N$. Because nonreciprocity is due to $\Delta\phi\ne0$, the nonreciprocal phase shift increases along this locus as well. Panel (b) shows that the frequency at which phase nonreciprocity occurs moves to lower values as the mass ratio increases. This can be understood by noting the increased mass of the system -- the non-dimensional masses of the unit cell are $1$ and $\mu$. 

\subsection{Influence of the forcing amplitude ($P$)}
\label{Fampl}

To investigate the influence of the forcing amplitude on phase nonreciprocity, we keep $\mu=2$ and compute the locus of $N^F=N^B$ as a function of $P$. Different loci are obtained depending on which of the diamond markers in Fig.~\ref{fig:4}(a) is being traced. We choose the intersection marked by the blue diamond for further investigation.

\begin{figure}
\begin{subfigure}[b]{.55\textwidth}
\centering
\includegraphics[width=\textwidth]{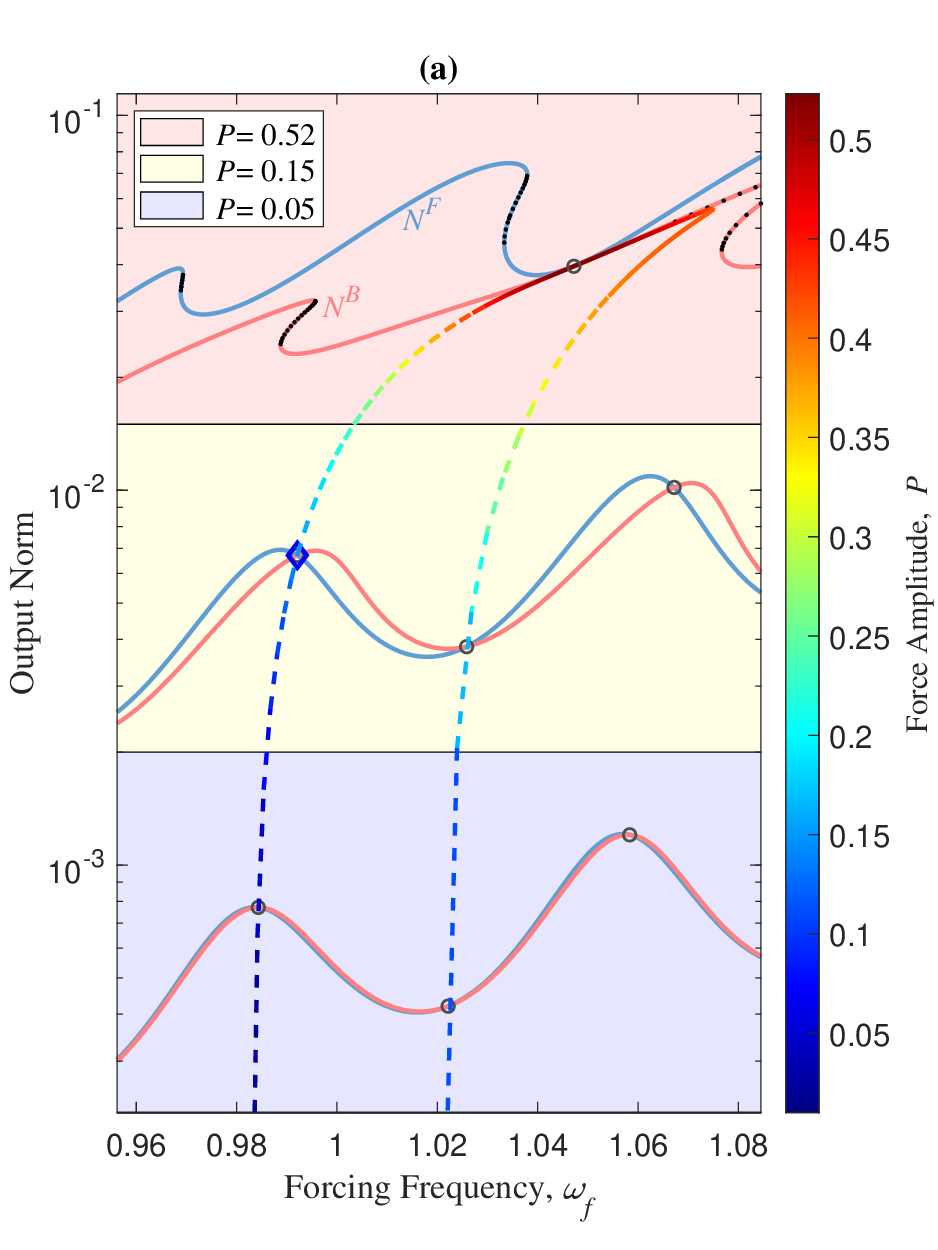}
\end{subfigure}
\begin{subfigure}[b]{.45\textwidth}
    \centering
    \begin{subfigure}[b]{\textwidth}
    \centering
    \includegraphics[width=\textwidth]{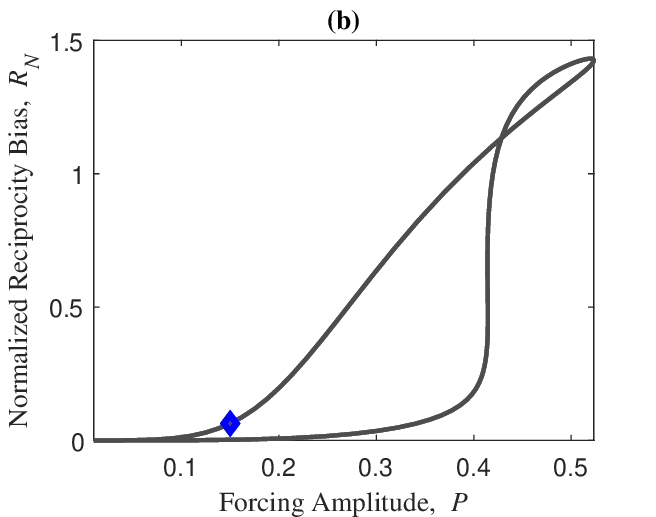}
    \end{subfigure}\\
    \begin{subfigure}[b]{\textwidth}
    \centering
    \includegraphics[width=\textwidth]{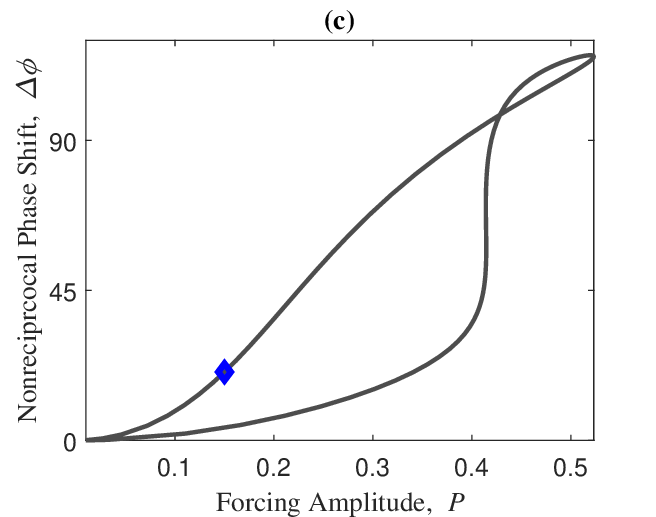}
    \end{subfigure}
    \end{subfigure}
    
    \caption{Locus of phase nonreciprocity ($N^F=N^B$) as a function of the forcing amplitude, $P$, for $\mu=2$. 
    (a)~Frequency response curves (solid) for three different values of $P$, superimposed onto backgrounds of different colors for clarity. The dashed, color-coded curve traces the locus of $N^F=N^B$; the color map corresponds to the value of $P$ along this curve. (b)~Normalized reciprocity bias, $R_N$. (c)~Nonreciprocal phase shift, $\Delta\phi$. 
    }
        \label{fig:6}
\end{figure}

Fig.~\ref{fig:6}(a) shows the frequency response curves at three values of $P$; these are superimposed onto backgrounds of different colors for clarity. The frequency response curves at $P=0.15$ are reproduced from Fig.~\ref{fig:2}. Starting from the intersection point at $\omega_f\approx0.99$ indicated by the blue diamond marker, the dashed curve traces the locus of $N^F=N^B$ as a function of $P$. Moving from the blue marker toward lower values of $P$, we see the trace reaching the frequency response curves at $P=0.05$. The intersection point is still near a resonance frequency of the system. 

Tracing the same locus toward increasing values of $P$, we reach an interaction point of the frequency response curves at $P=0.52$. We note here that the intersection is somewhat off-resonance for the forward configuration and near an anti-resonance for the backward configuration. The locus of phase nonreciprocity continues up to a maximum forcing amplitude of $P\approx0.5231$, above which the two frequency response curves do not intersect in this frequency range. Intersections at higher values of $P$ are not overruled, of course, but their discussion falls outside the scope of the present work. This is in part because we can no longer guarantee the harmonicity of the response; a different numerical algorithm would be required in this case~\cite{yousefzadeh_computation_2022}.

Moving along the locus of phase nonreciprocity beyond the turning point at $P\approx0.5231$, the value of $P$ decreases until we reach the intersection point with the second frequency response curves at $P=0.15$, near an anti-resonance of the system at $\omega_f\approx1.025$. The locus continues to a second intersection point with the frequency response curves at $P=0.05$, again close to an anti-resonance frequency of the system. 

We can now summarize the behavior of the locus of phase nonreciprocity for the intersection point indicated by the blue diamond. 
The locus starts at a resonance frequency at low forcing amplitudes and continues near the same resonance frequency as the value of $P$ increases. As the nonlinear effects become more prominent and the degree of nonreciprocity increases, the locus starts moving away from the resonance frequency and toward an anti-resonance frequency. The transition occurs through a turning point, after which the locus is near an anti-resonance frequency for both the forward and backward configurations. We observed a similar pattern for the other loci in the acoustic branch. 

Panels (b) and (c) in Fig.~\ref{fig:6} show the locus of phase nonreciprocity in the $(P,R_N)$ and $(P,\Delta\phi)$ planes, respectively. We see that the degree of nonreciprocity, as expected, is relatively higher along the portion of the locus that is near a resonance frequency of the system (higher response amplitude). The transition of the locus from the vicinity of a resonance frequency to the vicinity of an anti-resonance frequency is markedly evident in these panels by the sharp drop in $R_N$ just above $P=0.4$. Furthermore, the fact that $R_N$ follows $\Delta\phi$ so closely is a reminder that breaking of reciprocity is solely due to a nonreciprocal phase shift: $N^F=N^B$ while $\phi^F\ne\phi^B$. 

\begin{figure}
\begin{subfigure}[b]{0.55\textwidth}
\centering
\includegraphics[width=\textwidth]{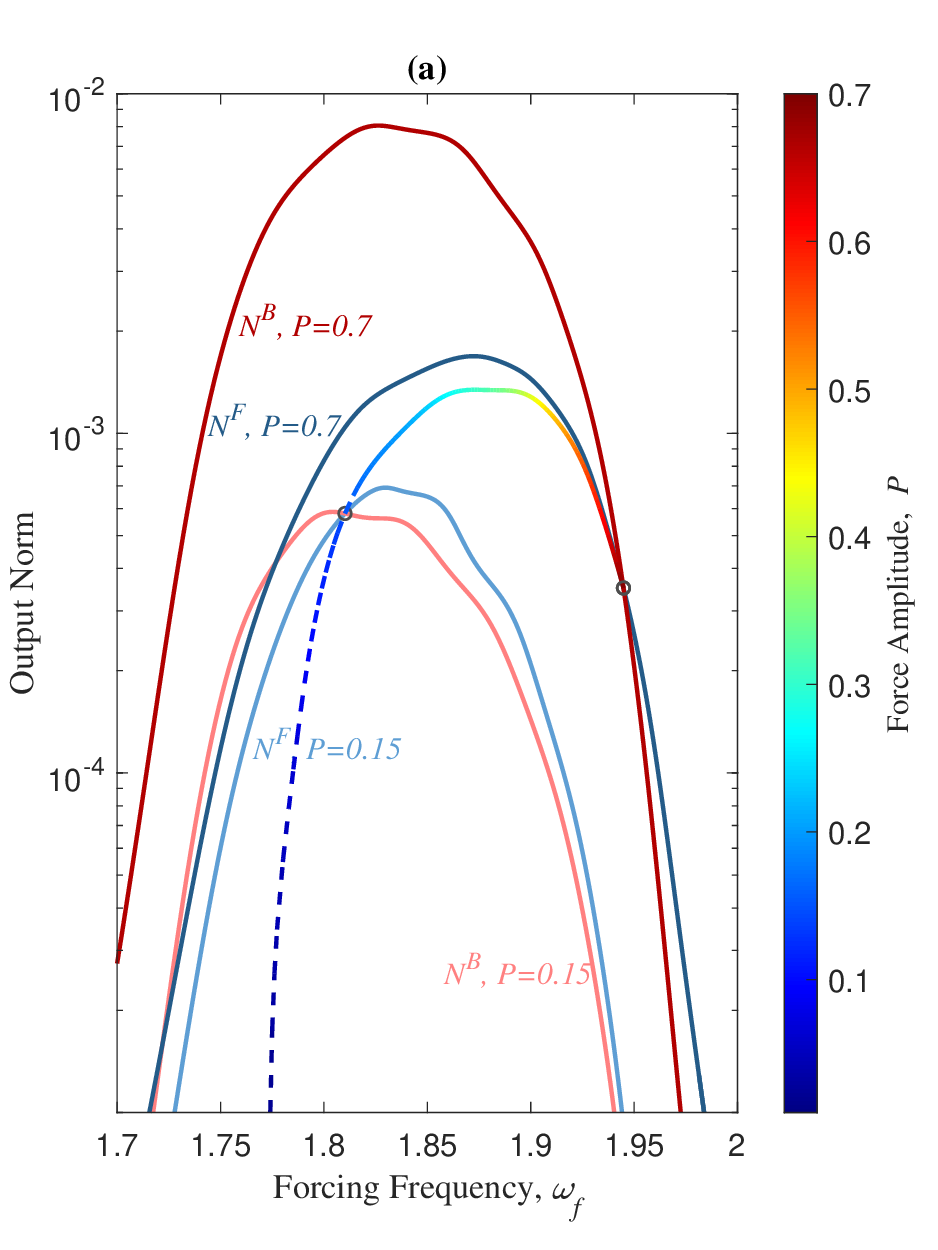}
\end{subfigure}
\begin{subfigure}[b]{0.45\textwidth}
    \centering
    \begin{subfigure}[b]{\textwidth}
    \centering
    \includegraphics[width=\textwidth]{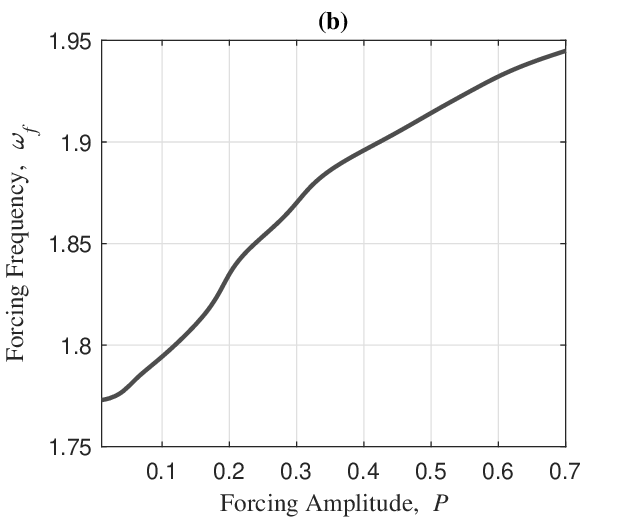}
    \end{subfigure}\\
    \begin{subfigure}[b]{\textwidth}
    \centering
    \includegraphics[width=\textwidth]{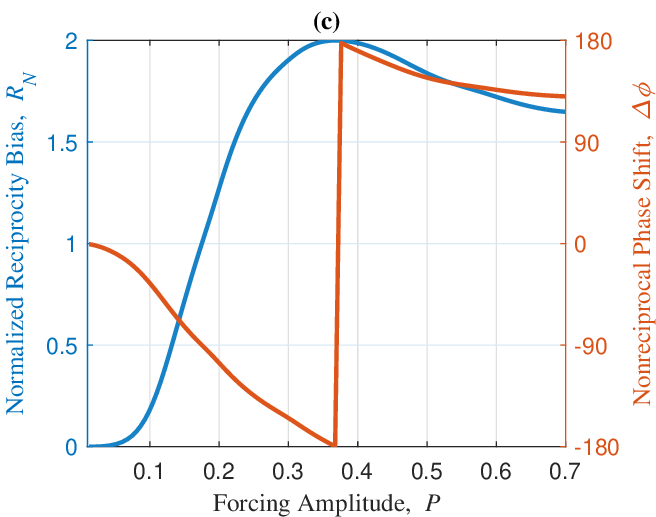}
    \end{subfigure}
    \end{subfigure}
    \caption{Locus of phase nonreciprocity ($N^F=N^B$) in the optical branch as a function of the forcing amplitude, $P$, for $\mu=2$. (a)~Frequency response curves for two different values of $P$. The dashed, color-coded curve traces the locus of $N^F=N^B$; the color map corresponds to the value of $P$ along this curve. 
    (b)~The forcing frequency, $\omega_f$, at which phase nonreciprocity occurs. 
    (c) Normalized reciprocity bias, $R_N$, and nonreciprocal phase shift, $\Delta\phi$, along the locus.}
        \label{fig:9}
\end{figure}

There is only one intersection of the frequency response curves at the optical branch near $\omega_f\approx1.8$. Fig.~\ref{fig:9}(a) shows the frequency response curves at two values of the forcing amplitude: $P=0.15$, which is the base configuration reproduced from Fig.~\ref{fig:3}, and $P=0.70$, which is chosen such that the response of the system remains harmonic and stable. The dashed curve shows the locus of $N^F=N^B$ for $0.01 \le P\le 0.7$. For higher values of $P$, the locus becomes unstable because the backward configuration loses stability through a Neimark-Sacker bifurcation; analysis of this operating regime falls outside the scope of this work. 

Moving along the locus of nonreciprocity Fig.~\ref{fig:9}(a), the output norm increases until $\omega_f\approx1.87$, after which it begins to decrease again. This decrease is accompanied by the intersection point moving away from the resonance frequency of the system, similar to what we observed for the acoustic branch. 
Fig.~\ref{fig:9}(b) shows the projection of the locus onto the ($P,\omega_f$) plane. This locus is somewhat `bumpy'. To explain this, we note that the modes within the optical branch are more damped than their acoustic counterparts; recall Fig.~\ref{fig:3}. Thus, as the forcing amplitude increases in Fig.~\ref{fig:9}(b), the locus traverses multiple modes of the system, which results in the observed behavior. The locus becomes relatively flat only after the locus has passed through the modes in the optical branch. This can be confirmed by observing the individual phases $\phi^F$
and $\phi^B$.

Fig.~\ref{fig:9}(c) shows that the normalized nonreciprocity bias, $R_N$, reaches a local maximum near $P\approx0.36$, which corresponds to $\omega_f\approx1.87$. This trend is mirrored in the nonreciprocal phase shift, $\Delta\phi$, because nonreciprocity is solely due to the phase difference. Accordingly, the maximum value of $R_N$ corresponds to $\Delta\phi=\pi$. The nonreciprocal phase shift, together with $R_N$, gradually decreases as the locus of phase nonreciprocity moves away from the resonant region of the forward and backward response curves. 

\subsection{Influence of damping ($\zeta_g$)}
\label{damping}

\begin{figure}
\begin{subfigure}[b]{0.55\textwidth}
\centering
\includegraphics[width=\textwidth]{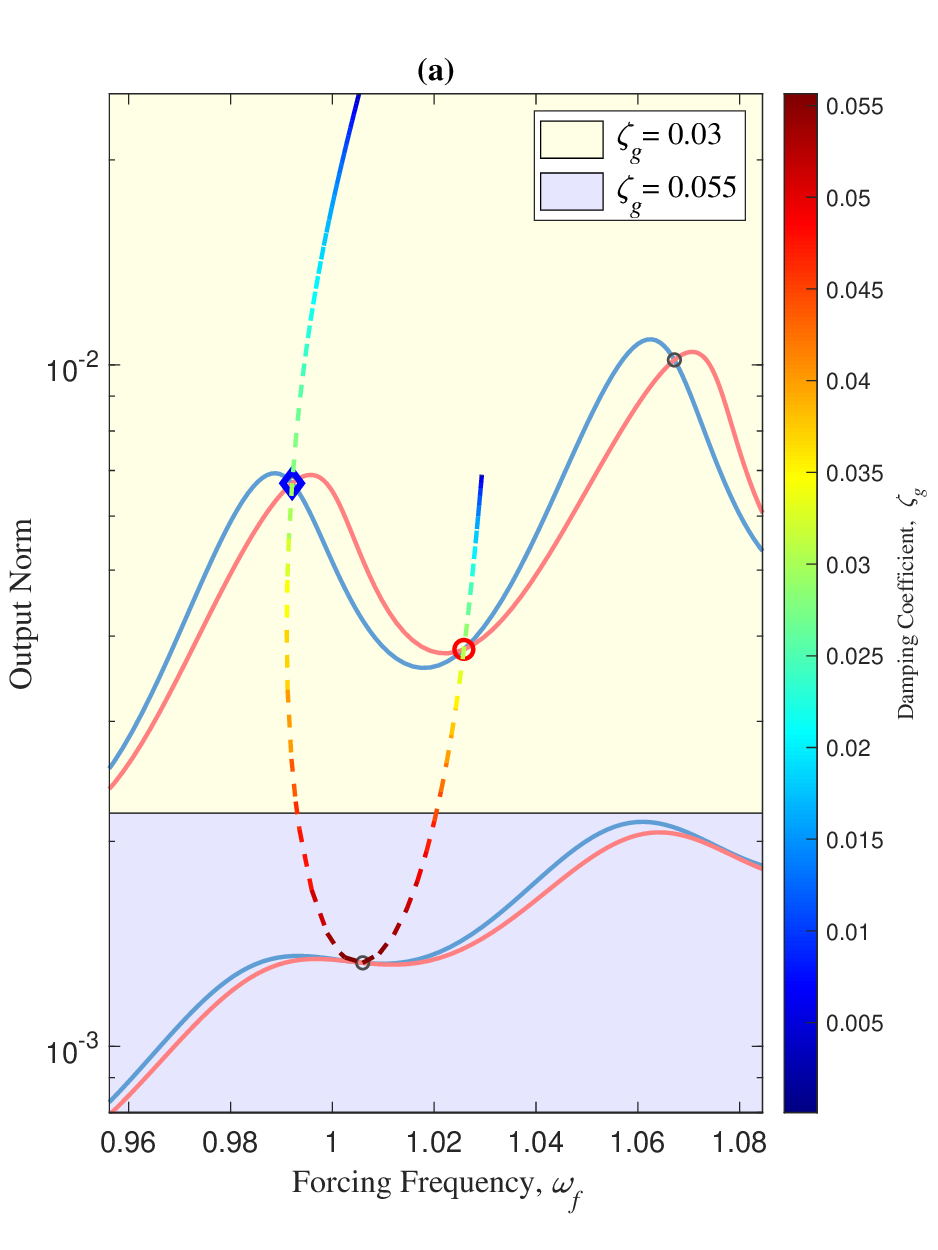}
\end{subfigure}
\begin{subfigure}[b]{0.45\textwidth}
    \centering
    \begin{subfigure}[b]{\textwidth}
    \centering
    \includegraphics[width=\textwidth]{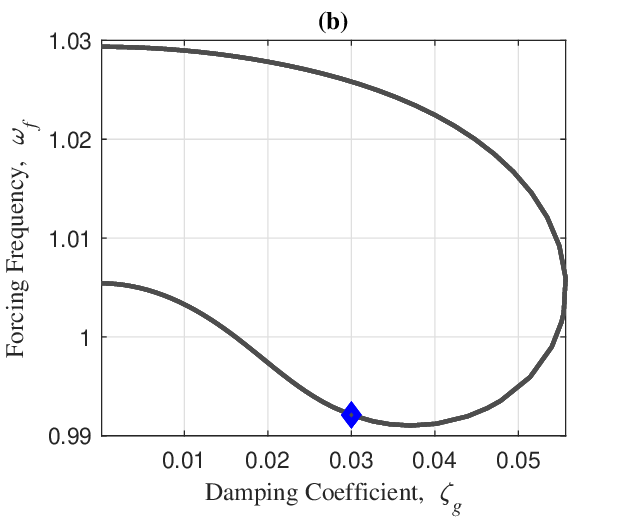}
    \end{subfigure}\\
    \begin{subfigure}[b]{\textwidth}
    \centering
    \includegraphics[width=\textwidth]{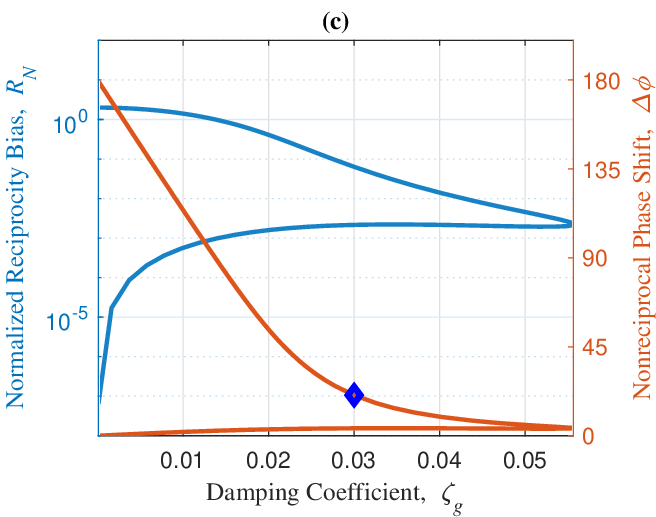}
    \end{subfigure}
    \end{subfigure}
    \caption{Locus of phase nonreciprocity ($N^F=N^B$) as a function of the damping ratio, $\zeta_g$, at $P=0.15$. (a)~Frequency response curves for two different values of damping. The dashed, color-coded curve traces the locus of $N^F=N^B$; the color map corresponds to the value of $\zeta_g$ along this curve. 
    (b)~The forcing frequency, $\omega_f$, at which phase nonreciprocity occurs. 
    (c) Normalized reciprocity bias, $R_N$, and nonrecirocal phase shift, $\Delta\phi$, along the locus. }
        \label{fig:7}
\end{figure}

Fig.~\ref{fig:7} shows the influence of damping on the locus of phase nonreciprocity at $P=0.15$. Panel~(a) shows the frequency response curves at two values of the damping ratio, $\zeta_g$, which are superimposed onto backgrounds of different colors for clarity. The frequency response curves at $P=0.15$ are reproduced from Fig.~\ref{fig:2}. We have traced the locus of $N^F=N^B$ (dashed curve) from two intersection points, at $\omega_f\approx0.99$ (blue diamond) and at $\omega_f\approx1.025$ (red circle); the color map corresponds to the value of $\zeta_g$ along this locus. 
Moving in the direction of increasing damping, we observe that the two loci merge together through a turning point at $\zeta_g\approx0.055$; panel~(b) shows the projection of the locus onto the $(\zeta_g,\omega_f)$ plane. We observe in the frequency response curves in panel~(a) that, as $\zeta_g$ increases, the influence of nonlinear forces is becoming increasing smaller, making the response more linear gradually. 
We recall that the overall effect of increasing damping is, indeed, to merge the two frequency response curves because it diminishes the relative effect of nonlinearity and, colloquially speaking, linearizes the response of the system. 

Fig.~\ref{fig:7}~(c) shows the evolution of the normalized reciprocity norm, $R_N$, and the nonreciprocal phase shift, $\Delta\phi$, along the locus of phase nonreciprocity. We note that as the value of $\zeta_g$ increases, both $R_N$ and $\Delta\phi$ diminish rapidly until the turning point. As the value of $\zeta_g$ decreases toward zero, the locus of phase nonreciprocity terminates either at $(\Delta\phi,R_N)=(0,0)$ or at $(\Delta\phi,R_N)=(\pi,2)$. The end point with $\Delta\phi=0$ corresponds to a matching of the anti-resonance frequencies of the forward and backward configurations, while $\Delta\phi=\pi$ corresponds to a matching of their resonances.

\begin{figure}[h!]
\begin{subfigure}{1\textwidth}
  \includegraphics[width=\textwidth]{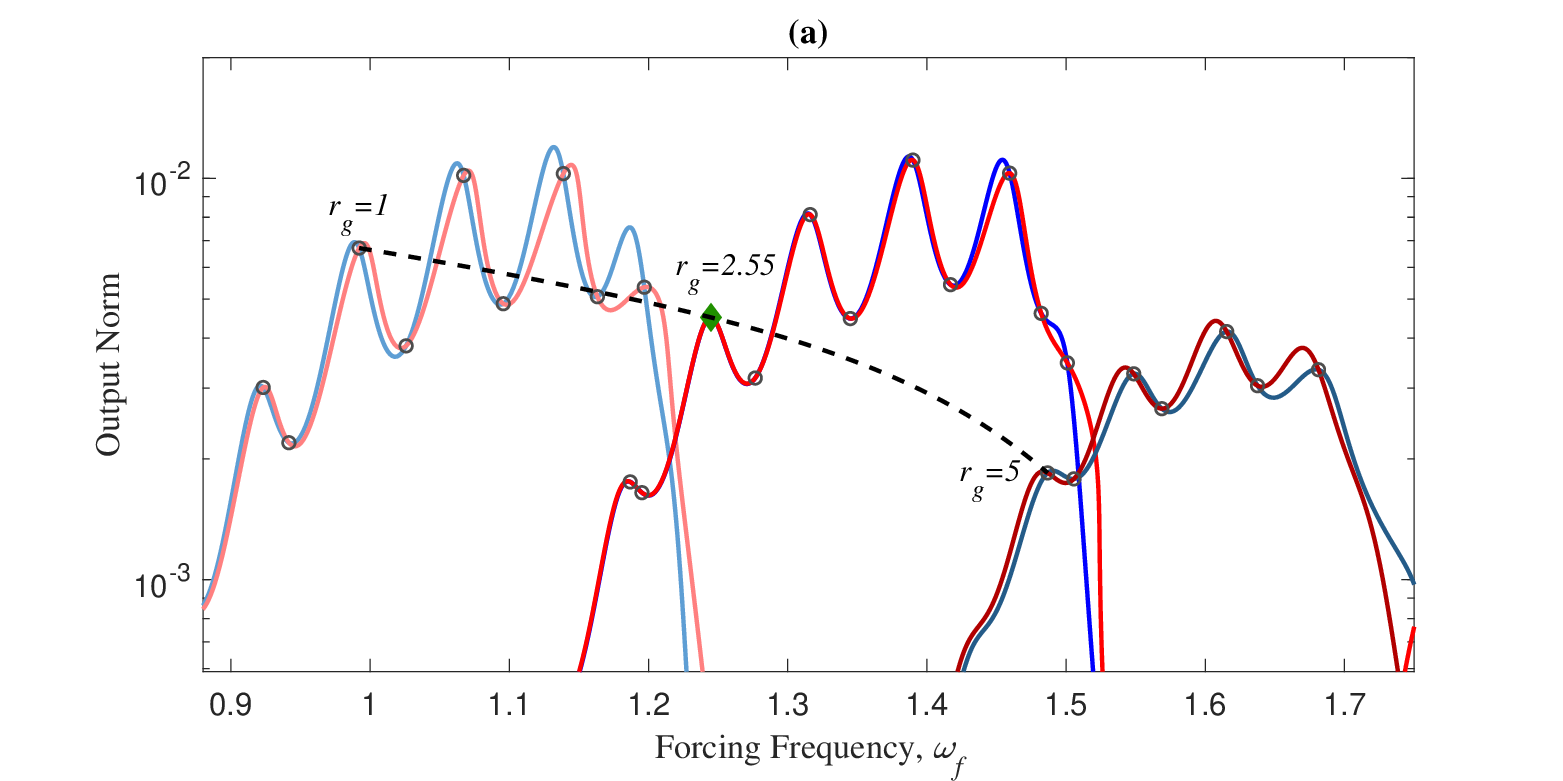}
\end{subfigure}

\begin{subfigure}[b]{.49\textwidth}
\includegraphics[width=\textwidth]{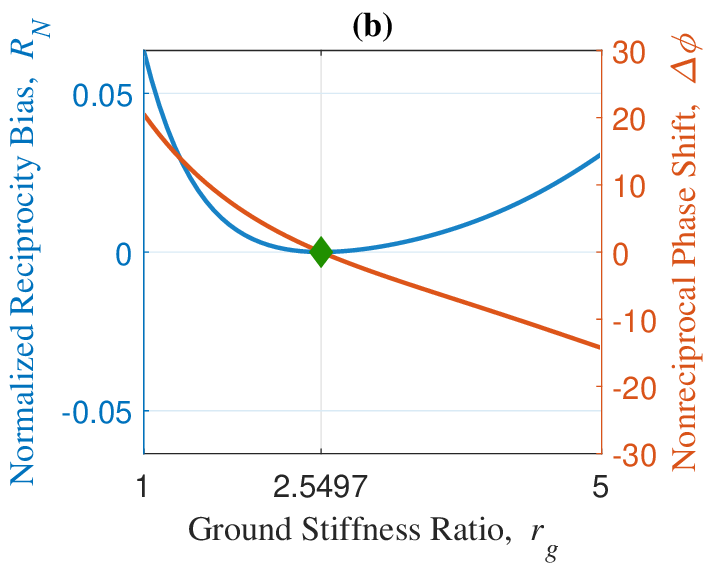}
\end{subfigure}
\hfill
\begin{subfigure}[b]{.49\textwidth}
\includegraphics[width=\textwidth]{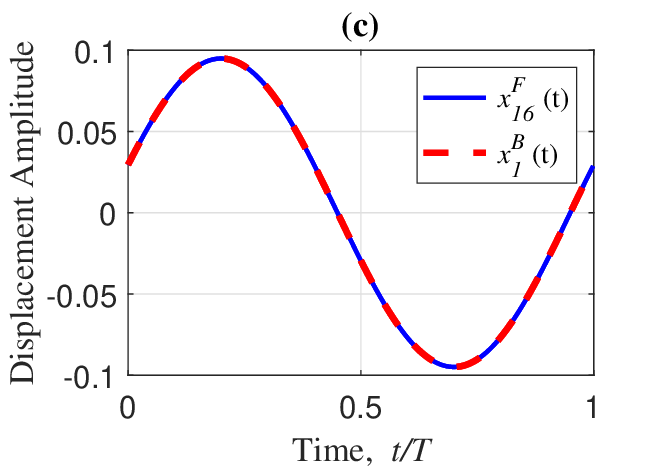}
\end{subfigure}
    \caption{
    Restoring reciprocity $R_N=0$ for $(P,\mu)=(0.15,2)$ by tuning a second symmetry-breaking parameter, $r_g$. 
    (a)~Output norms for three different values of the stiffness ratio. The black dashed curve shows the locus of $N^F=N^B$. The green diamond marker denotes the point at which reciprocity is restored. 
    (b)~Variation of $R_N$ and $\Delta\phi$ as a function of $r_g$. The green diamond marker denotes the point at which reciprocity is restored. 
    (c)~Time-domain response of the forward and backward configurations at the green diamond marker ($R_N=0$). }
  
\label{fig:8}
\end{figure}

\section{Restoring reciprocity using two symmetry-breaking parameters}
\label{restoring}

The forward and backward configurations have equal output norms for phase nonreciprocity, $N^F=N^B$. If, additionally, the nonreciprocal phase shift can be set to zero, $\Delta\phi=0$, then the output displacements are identical and we retrieve a reciprocal response, $R=0$. To achieve this, a second symmetry-breaking parameter (other than $\mu$) is required to counterbalance the effect of the existing asymmetry and restore reciprocity~\cite{giraldo_restoring_2023}. The two symmetry-breaking parameters thus act together to maintain reciprocity in a system with broken mirror symmetry. We use the stiffness ratio, $r_g$, as the second symmetry-breaking parameter. From a practical perspective, we note that it is possible to independently tune the effective mass and stiffness parameters of a mechanical system~\cite{Frazier}.

Fig.~\ref{fig:8}(a) shows frequency response curves for the system with $\mu=2$ at $P=0.15$ at three different values of $r_g$. The frequency response curves on the left side correspond to $r_g=1$, where $\mu$ is the only symmetry-breaking parameter. The dashed black curve traces the locus of phase nonreciprocity that emanates from the intersection point near $\omega_f\approx0.99$ as the value of $r_g$ increases. 

Fig.~\ref{fig:8}(b) shows the corresponding values of the normalized reciprocity norm, $R_N$, and nonreciprocal phase shift, $\Delta\phi$, along the locus of phase nonreciprocity. We note that the value of $R_N$ becomes zero near $r_g\approx2.55$, indicating a reciprocal response. Because $R_N$ is a non-negative number, we use $\Delta\phi$ to confirm whether the response is indeed reciprocal. The nonreciprocal phase shift has a zero-crossing near $r_g\approx2.55$, which confirms reciprocity. Fig.~\ref{fig:8}(c) shows the corresponding reciprocal response in the time domain. 

We return to Fig.~\ref{fig:8}(a) to observe the frequency response curves at $r_g\approx2.55$. We observe that the forward and backward configurations have similar norms in the vicinity of the resonance frequency where reciprocity is restored; {\it i.e.}, near the intersection point at $\omega_f\approx1.24$. Similar findings are reported in a system of coupled oscillators~\cite{giraldo_restoring_2023}. When $r_g$ is further increased, the two frequency response curves drift apart again. Phase nonreciprocity persists, but the response is no longer reciprocal ($\Delta\phi\ne0$).

\section{Conclusion}
\label{conclusion}
We presented a computational analysis of nonreciprocal vibration transmission in a discrete model of a nonlinear periodic material. Nonlinearity and asymmetry are the two required ingredients for realizing nonreciprocity in this setting. Nonlinearity appeared in the grounding elasticity as a cubic spring (symmetric restoring force). Asymmetry appeared at the substructure level in two ways: the ratios of the effective inertia and effective linear elasticity of the two units. We focused on the weakly nonlinear steady-state response of the system to a harmonic excitation; {\it i.e.}, the frequency-preserving regime. We presented scenarios in which there is a phase difference between the transmitted vibrations in the forward and backward configurations ($\phi^F\ne\phi^B$), but the energies transmitted in the opposite directions are equal ($N^F=N^B$). Thus, nonreciprocity is due {\it solely} to the nonreciprocal phase shifts ($\Delta\phi=\phi^F-\phi^B$) in the transmitted vibrations. We call this the state of {\it phase nonreciprocity}. 

We performed a parametric study of phase nonreciprocity in the nonlinear waveguide, investigating the influence of the forcing amplitude, damping ratio, and degree of asymmetry on the nonreciprocal phase shifts. In particular, by tuning two independent symmetry-breaking parameters (mass ratio and stiffness ratio), we have shown that it is possible to restore reciprocity in the vicinity of a resonance frequency of the system. We achieved this by finding a nontrivial set of parameters at which $\Delta\phi=0$ along the locus of phase nonreciprocity. The results indicate that although breaking the symmetry of the system is a necessary requirement for enabling nonreciprocal dynamics, it is not strictly a sufficient condition. Furthermore, these results showcase the potential of asymmetry to serve as an additional design parameter, especially in systems that rely on symmetry but are inherently asymmetric due to unavoidable imperfections. 
We hope that these findings contribute to enhancing the performance of devices that operate based on nonlinear nonreciprocity.

\section*{Acknowledgments}
We acknowledge financial support from the Natural Sciences and Engineering Research Council of Canada through the Discovery Grant program. A.K. acknowledges additional support from Concordia University. 

\nocite{*}

\section*{Conflict of Interest}

The authors declare that they have no conflict of interest.

\section*{Data Availability}

The datasets generated during this study can be made available from the corresponding author on reasonable request.

\bibliographystyle{unsrt} 
\bibliography{mypaper.bib}

\section*{Appendix A: Non-dimensional Equations of Motion}
\label{appA}

The governing equations for the system in Fig.~\ref{fig:1} can be written as:
\begin{equation}
\begin{aligned}
\label{EOMraw}
M_1\ddot{x}_{2i-1}+2k_cx_{2i-1}-k_c(x_{2i-2}+x_{2i})+k_gx_{2i-1}+k_nx_{2i-1}^3+c\dot{x}_{2i-1}=f_{2i-1}\cos{\omega_ft} \\
M_2\ddot{x}_{2i}+2k_cx_{2i}-k_c(x_{2i+1}+x_{2i-1})+k'_gx_{2i}+k_nx_{2i}^3+c\dot{x}_{2i}=f_{2i}\cos{\omega_ft}
\end{aligned}
\end{equation}
where $i=1,...,8$ is the counter of the unit cells, $k_c$ is the coupling stiffness, $k_g$ and $k'_g$ are the coefficients of the linear grounding springs for $M_1$ and $M_2$, and $c$ is the linear viscous damping connecting each mass to the ground. We divide the equations by $k_g$ and introduce the non-dimensional parameters $\tau=\omega_0t$, $\omega_0^2=k_g/M_1$, $\Omega=\omega_f/\omega_0$ to obtain 
\begin{equation}
\begin{aligned}
M_1\omega_0^2/k_gx''_{2i-1}+2k_c/k_gx_{2i-1}-k_c/k_g(x_{2i-2}+x_{2i})+x_{2i-1}+k_n/k_gx_{2i-1}^3+2\zeta_gx'_{2i-1}=f_{2i-1}/k_g\cos{\Omega\tau} \\
M_2\omega_0^2/k_gx''_{2i}+2k_c/k_gx_{2i}-k_c/k_g(x_{2i+1}+x_{2i-1})+k'_g/k_gx_{2i}+k_n/k_gx_{2i}^3+2\zeta_gx'_{2i}=f_{2i}/k_g\cos{\Omega\tau}
\end{aligned}
\end{equation}
where $x'=dx/d\tau=(dx/dt)/\omega_0$, $x''=d^2x/d\tau^2=(d^2x/dt^2)/\omega_0^2$ and $\zeta_g=(c\omega_0)/(2k_g)$. 
We define the non-dimensional displacement and force as $\bar{x}=x/d$ and $F=f/(dk_g)$, where $d$ is a characteristic displacement of the system. This results in 
\begin{equation}
\label{eqNonD}
\begin{aligned}
\bar{x}''_{2i-1}+2K_c\bar{x}_{2i-1}-K_c(\bar{x}_{2i-2}+\bar{x}_{2i})+\bar{x}_{2i-1}+K_n\bar{x}_{2i-1}^3+2\zeta_g\bar{x}'_{2i-1}=F_{2i-1}\cos{\Omega\tau} \\
\mu\bar{x}''_{2i}+2K_c\bar{x}_{2i}-K_c(\bar{x}_{2i+1}+\bar{x}_{2i-1})+r_g\bar{x}_{2i}+K_n\bar{x}_{2i}^3+2\zeta_g\bar{x}'_{2i}=F_{2i}\cos{\Omega\tau}
\end{aligned}
\end{equation}
where $\mu=M_2/M_1$, $K_c=k_c/k_g$, $K_n=d^2k_c/k_g$ and $r_g=k'_g/k_g$. 
Eq.~(\ref{eqNonD}) is the non-dimensional form of Eq.~(\ref{EOMraw}). Eq.~(\ref{eqNonD}) is the same as Eq.~(\ref{govern}) in the main text, where we have dropped the overbar in $\bar{x}$ and replaced $\tau$ with $t$ for ease of reference.
\end{document}